\documentclass[fleqn,usenatbib]{mnras}

\usepackage{graphicx}	
\usepackage{amsmath}	
\usepackage{amssymb}	
\usepackage{tikz}
\title[Discovery of mHz QPOs in the X-ray binary EXO 0748$-$676]{Discovery of millihertz quasi-periodic oscillations in the X-ray binary EXO 0748$-$676}

\author[G. C. Mancuso et al. ]{G. C. Mancuso,$^{1,2,3}$\thanks{E-mail: gmancuso@iar.unlp.edu.ar} D. Altamirano,$^{3}$ F. Garc\'{\i}a,$^{1,4}$ M. Lyu,$^{5}$ M. M\'{e}ndez,$^{6}$  
\newauthor
J. A. Combi,$^{1,2}$ M. D\'{\i}az-Trigo$^{7}$ and J. J. M. in't Zand$^{8}$\\
$^{1}$Instituto Argentino de Radioastronom\'{\i}a (CCT-La Plata, CONICET; CICPBA), C.C. No. 5, 1894 Villa Elisa, Argentina\\
$^{2}$Facultad de Ciencias Astron\'omicas y Geof\'{\i}sicas, Universidad Nacional de La Plata, Paseo del Bosque s/n, 1900 La Plata, Argentina\\
$^{3}$Physics \& Astronomy, University of Southampton, Southampton, Hampshire SO17 1BJ, UK\\
$^{4}$Laboratoire AIM, CEA, CNRS, Universit\'e Paris-Saclay, Universit\'e Paris Diderot, Sorbonne Paris Cit\'e, F-91191 Gif-sur-Yvette, France\\
$^{5}$Department of Physics, Xiangtan University, Xiangtan, Hunan 411105, China\\
$^{6}$Kapteyn Astronomical Institute, University of Groningen, PO BOX 800, NL-9700 AV Groningen, the Netherlands\\
$^{7}$ESO, Karl-Schwarzschild-Strasse 2, 85748 Garching bei M{\"u}nchen, Germany\\
$^{8}$SRON Netherlands Institute for Space Research, Sorbonnelaan 2, 3584 CA Utrecht, The Netherlands}

\begin{document}
%
%
%
\maketitle
\label{firstpage}
\begin{abstract}
We report the discovery of millihertz quasi-periodic oscillations (mHz QPOs) from the bursting, high-inclination atoll neutron-star low-mass X-ray binary (NS LMXB) EXO 0748--676 with the Rossi X-ray Time Explorer (RXTE). This class of QPO, originally discovered in three NS LMXBs, has been interpreted as a consequence of a special mode of nuclear burning on the NS surface. Using all the RXTE archival observations of the source, we detected significant ($>3\sigma$) mHz QPOs in 11 observations. The frequency of the oscillations was between $\sim$ 5 and $\sim$ 13 mHz. We also found a decrease of the QPO frequency with time in two occasions; in one of these the oscillations disappeared with the onset of an X-ray burst, similar to what was reported in other sources. Our analysis of the X-ray colours revealed that EXO 0748--676 was in a soft spectral state when it exhibited the QPOs. This makes EXO 0748--676 the sixth source with mHz oscillations associated to marginally stable burning, and the second one that shows a systematic frequency drift. Our results suggest that the mechanism that produces the drift might always be present if the mHz QPOs are observed in the so-called intermediate state.

\end{abstract}


%
%
%
%

\section{Introduction}\label{sec:intro}

Low-mass X-ray binaries (LMXBs) are some of the brightest X-ray sources in the Galaxy. These systems consist of a compact object (a neutron star or a black hole) and a `normal' (generally a main sequence $\lesssim$~1\hspace{.7mm}$M_{\odot}$) companion star. These LMXBs are powered by mass accretion, from the donor star to the compact object, through Roche-lobe overflow. Because of the conservation of angular momentum, this material does not fall directly onto the compact star but forms an accretion disk around it \citep{pringle1972}.

NS LMXBs are classified into two main groups based on their timing and spectral properties, the Z and atoll sources, being the latter fainter than the former \citep{hasinger1989}. In turn, the atoll sources are divided in three main spectral states: hard (or extreme island), soft (or banana) and intermediate (or island) states.
NSs in LMXBs are usually thought to be weakly magnetized (B $<$ 10$^{10}$\hspace{1mm}G; e.g., \citealt{psaltis1999} and references therein). Many of these systems show X-ray bursts, a sudden and very rapid increase in the X-ray flux due to a thermonuclear runaway, followed by an exponential decay due to the cooling of the NS \citep[see, e.g.,][]{strohmayer2006}. Most of the NS LMXBs show X-ray variability in the form of relatively sharp mHz to kHz quasi-periodic oscillations (QPOs), and broad band noise at $<100$ Hz (see, e.g., review by \citealt{vanderklis2006}).
Typical well defined classes of QPOs are, among others, the so-called
(i)~kHz QPOs, which generally appear in pairs in the 300--1300 Hz range (see, e.g., \citealt{vanderklis2006}), 
(ii) hectohertz (hHz) QPOs, with frequencies in the range of $\sim$ 100--300 Hz (e.g., \citealt{vanstraaten2003}, \citealt{altamirano2008a}),
%
(iii) low-frequency QPOs, which have frequencies in the 0.01--50 Hz (e.g., \citealt{jonker1999,jonker2000}; \citealt{homan2003}; \citealt{vanstraaten2003}; \citealt{wijnands2004}) depending on the source state and luminosity, and 
(iv) mHz QPOs, observed at frequencies between 6 and 15 mHz in five NS LMXBs (4U 1636--53, 4U 1608--52 and Aql X--1, \citealt{revnivtsev2001};  4U 1323--619, \citealt{strohmayer2011}; and GS 1826--238, \citealt{strohmayer2018}),\footnote{\citet{linares2010} reported mHz QPOs from the 11 Hz pulsar IGR J17480--2446, but the properties of the QPOs are quite different compared with those of the listed sources. \citet{ferrigno2017} found mHz QPOs in the accreting millisecond X-ray pulsar IGR J00291+5934, but the QPO origin is unclear.} all of which are reported or candidate atoll sources.  


The mHz QPOs have properties that set them apart from the other types of QPOs: mHz QPOs occur only within a narrow range of X-ray luminosities ($L_{2-20\hspace{0.7mm} \rm keV} \simeq (0.5-1.5) \times 10^{37}$ erg s$^{-1}$), their fractional rms amplitude increases towards low energies, and the QPOs disappear at the onset of a thermonuclear (type I) X-ray burst (\citealt{revnivtsev2001}, \citealt{altamirano2008b}). 
\citet{revnivtsev2001} suggested that the mHz QPOs could correspond to a special mode of nuclear burning on the NS surface, although they could not rule out an explanation due to instabilities in the accretion disc. The results of \citet{Yu2002} further supported the thermonuclear interpretation: they found that in the atoll source 4U 1608--52 the frequency of the kHz QPO was anti-correlated with the 2--5 keV X-ray count rate variations during the mHz oscillation. This anti-correlation suggests that the inner edge of the accretion disc increases due to the additional flux produced in each mHz QPO cycle on the NS surface. This is opposite to the positive correlation observed on longer timescales between the kHz QPO frequency and the X-ray count rate, consistent with the radius of the inner disc edge decreasing as the accretion rate increases \citep[see, e.g.,][]{mendez1999}.

\citet{altamirano2008b} found a systematical decrease of the mHz QPO frequency with time in the NS LMXB 4U 1636--53. Once the frequency dropped below 9 mHz, a bright type I X-ray burst occurred and the oscillations disappeared. This drifting behaviour was only observed when 4U~1636--53 was close to the transition between the hard and soft states. \citet{altamirano2008b} showed that the mHz QPO frequency can be used to predict the occurrence of a type~I burst, strengthening the interpretation that the mHz QPO phenomenon is intimately related to nuclear burning on the NS surface.

Additional work of \citet{lyu2014,lyu2015,lyu2016} investigated the properties of the mHz QPOs in 4U 1636--53. Using RXTE and XMM-Newton observations, the authors found that the mHz QPO frequency and the temperature of the NS surface were not significantly correlated. This finding is not consistent with theory, which predicts an anti-correlation between those parameters (\citealt{heger2007}). These authors also found that all the bursts that come up immediately after the mHz QPOs had positive convexity, suggesting that both phenomena happen at the NS equator.

From the theoretical side, \citet{heger2007} suggested that the mHz QPOs are produced by marginally stable nuclear burning of He on the NS surface. The model of \citet{heger2007} predicts an oscillatory mode of burning, with a period of $\approx100$ seconds (similar to what is observed), and only happening near the transition between stable and unstable burning (first predicted by \citealt{paczynski1983}), i.e., within a very narrow range of mass accretion rate. 
However, while in this model the oscillations appear at accretion rates close to the Eddington rate, observationally the QPOs were observed at an order of magnitude lower $L_{\rm X}$. \citet{heger2007} explained the discrepancy by proposing that the accreted material covers only $\sim10$\% of the NS surface, so that the local accretion rate (accretion rate per unit area) is effectively close to that of Eddington.

The model of \citet{heger2007}  cannot explain all the phenomenology observed (e.g. mHz QPOs and X-ray bursts occurring at the same time and the frequency drift). To further investigate this issue, \citet{keek2014} studied the possible relation between the mHz QPOs, the composition of the accreted fuel and the nuclear reaction rates. They found that neither the variation in the composition nor the reaction rate are responsible for the appearance of the mHz QPOs at the observed accretion rate. However, \citet{keek2009} investigated the effect of NS rotation on the stability of the burning. They found that the combination of a turbulent chemical mixing of the accreted fuel and a higher heat flux from the crust increase the stability of He burning, such that it could explain the observed accretion rate at which the mHz QPOs take place. 
Most importantly, \citet{keek2009} found that the QPO frequency correlates with the heat flux from the crust, so that the frequency drift observed by \citet{altamirano2008b} could be due to the cooling of the deeper layers of the NS where the burning occurs. 


EXO 0748--676 is a transient NS LMXB 
discovered in outburst during EXOSAT observations in February of 1985 (\citealt{parmar1985}). Its X-ray light curve exhibits $\sim 8.3$-min duration eclipses, implying an inclination of $\sim 75\textsuperscript{o} - 82\textsuperscript{o}$ (\citealt{parmar1986}). The eclipses allowed to estimate an orbital period of $\sim 3.82$ hr. The source also shows irregular X-ray dips, mostly in the half orbital cycle prior to the eclipse (\citealt{parmar1986}). EXO 0748--676 also presents type~I X-ray bursts, a few of which exhibited photospheric radius expansion (PRE). Depending on the assumed fuel composition, the distance estimated using PRE bursts was between $d=5.9 \pm 0.9$~kpc and $7.4 \pm 0.9$~kpc (\citealt{wolff2005}; \citealt{galloway2008}). \citet{homan2000} detected a kHz QPO during the persistent emission of an observation of this source with RXTE. \citet{galloway2010} reported a 552-Hz burst oscillation, associated with the spin frequency of the NS in EXO 0748--676. 
%

\begin{figure}
\includegraphics[height=6.1cm, width=\columnwidth]{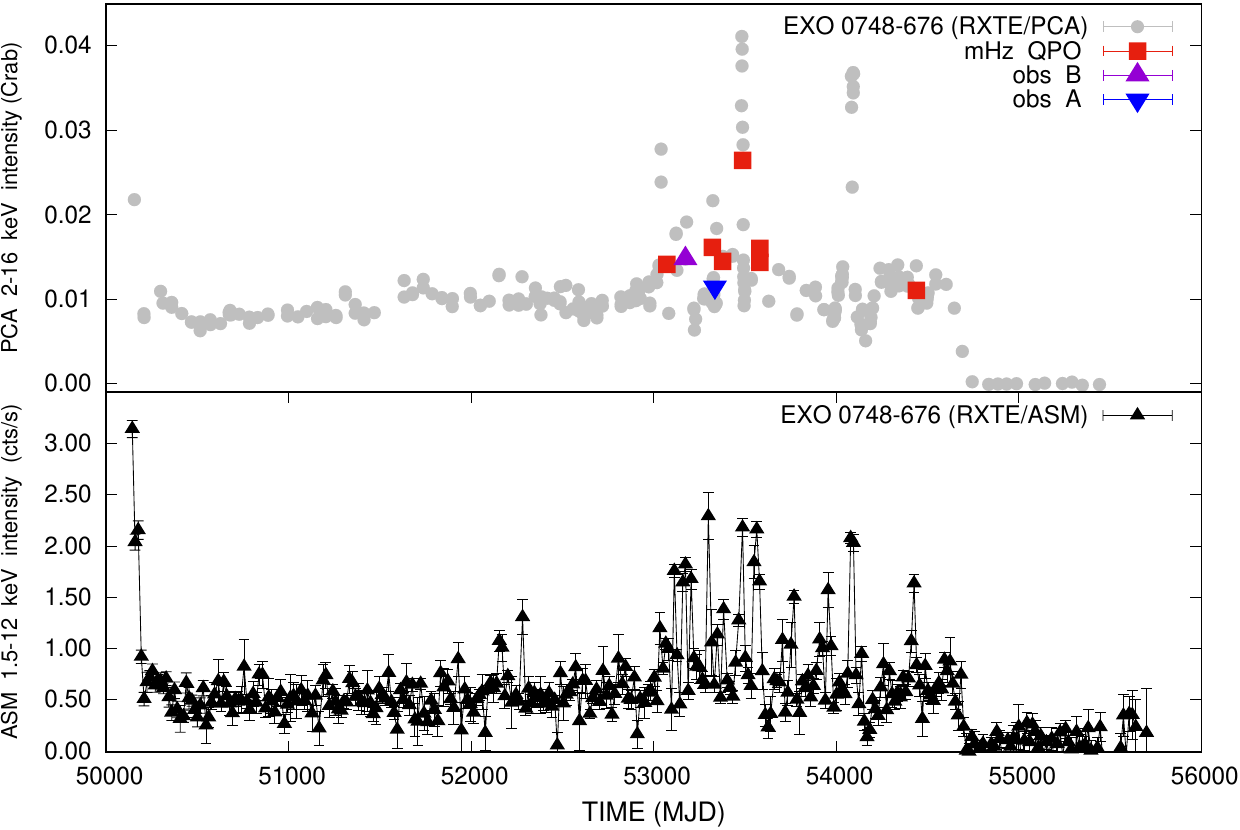}
\vspace{-0.6cm}
\caption{\emph{Upper panel:} 1996--2010 RXTE/PCA (2--16 keV, filled grey circles) light curve of EXO 0748--676. Each point corresponds to the average intensity per day when data were available (all type I bursts and eclipses have been removed). The filled red squares represent the daily-average count rate of the source when the mHz QPOs were found. Note that these cases are concentrated mostly in a specific period of time. Note also that the source faded from around MJD $\sim$ 54800. \emph{Lower panel:} 1996--2010 15-day averaged RXTE/ASM (1.5--12 keV, filled black triangles) light curve of EXO 0748--676. Several flares were detected from MJD $\sim$ 53000, until the source transitioned to the quiescent state (MJD $\sim$ 54800).}
\label{fig:LCcomplete}
\end{figure}

%
%
%
%

\vspace{-0.6cm}
\section{Observations and data analysis}\label{sec:dataanalysis}

We used all pointed observations of EXO 0748$-$676 obtained with the Proportional Counter Array (PCA; for instrument information see \citealt{jahoda2006}) aboard the Rossi X-ray Timing Explorer (RXTE). We studied a total of 749 observations taken over nearly fifteen years (March 1996 -- September 2010). An observation is usually composed of more than an orbit (each one of $\sim$ 90 minutes long duration), separated by data gaps of at least 1 ksec. In a few occasions the conditions were such that RXTE continuously observed the source for up to 17 ksec. Thus, our datasets consist of data segments of length $\sim$ 0.4 -- 17 ksec. 
We used 1 s resolution PCA (observed in Event, Good-Xenon, and Single Bit modes) data to create  light curves. Under the assumption that the mHz QPOs are stronger at lower energies, we followed \citet{altamirano2008b} and searched for them in the $\approx$ 2--5 keV energy range (channels 0--10), applying the Lomb-Scargle periodogram (\citealt{lomb1976}; \citealt{scargle1982}; \citealt{press1992}) to each gap-free segment separately. 
In those segments in which one or more type I X-ray bursts were detected, we searched for periodicities before, after and in between bursts. When an eclipse was detected, we searched for variability before and after it. We found that the oscillations in the $\approx$ 2--5 keV energy band are evident from the light curves. In all cases, the detections reported in this {\it Letter} are above the 3$\sigma$ level. Following \citet{altamirano2008b}, we calculate the significance level taking into account the number of frequencies searched and assuming white noise as prescribed in the method of \citet{press1992}. 
We also measured the frequency of the mHz QPO, $\nu_{\rm QPO}$, as the frequency at which the power reaches its maximum value in its respective periodogram.

In order to create the colour-colour and hardness-intensity diagrams, we calculated colours using the 16-s time-resolution Standard 2 mode data and followed the exact same method used by \citet{altamirano2008a}, with the addition that for this source we identified and excluded eclipsing episodes as well as type I X-ray bursts. We defined the soft colour as the 3.5--6.0 keV / 2.0--3.5 keV count rate ratio, the hard colour as the 9.7--16.0 keV / 6.0--9.7 keV count rate ratio, and the intensity as the 2.0--16.0 keV count rate. All values were normalized by the colours of the Crab in the same energy bands.

\begin{figure}
\includegraphics[height=6.1cm, width=0.8\columnwidth]{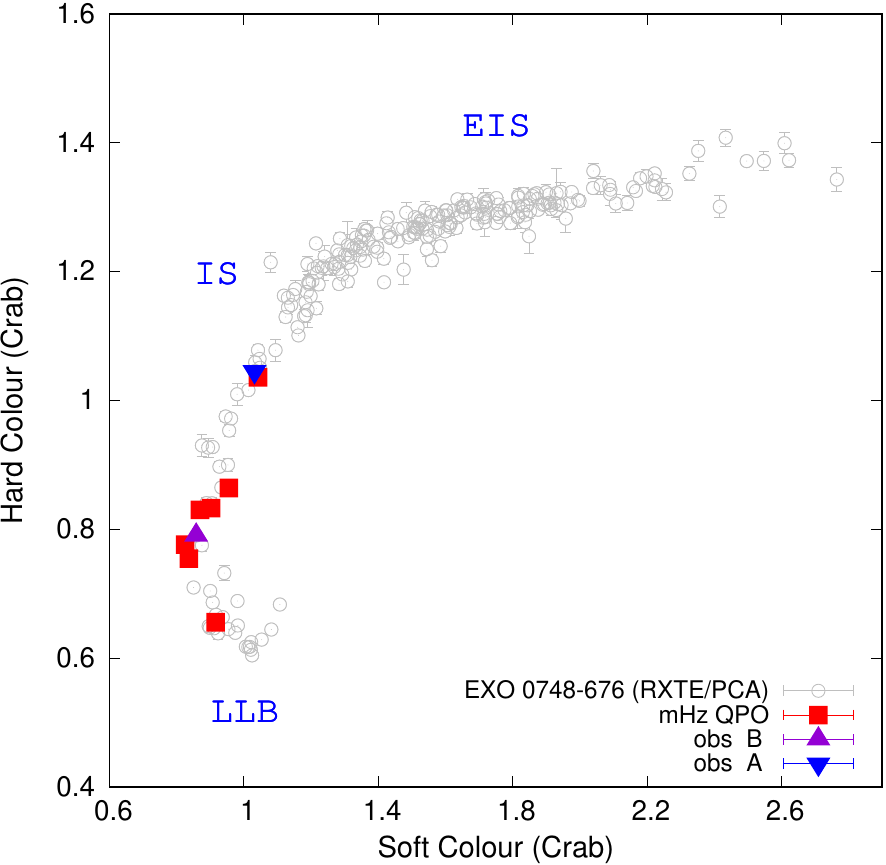}
\centering
\caption{Colour-colour diagram (CCD) of all RXTE observations of EXO 0748--676. Each open grey circle represents the daily-average colour of the source from all RXTE observations, excluding the times of bursts, eclipses, and dips. The colours of EXO 0748--676 are normalized to the colours of Crab. The filled red squares represent the colour of the persistent emission of the source when the mHz QPOs were found. Note also that the source was only observed in certain states: extreme island state (EIS), island state (IS) and lower left banana (LLB).}
\label{fig:CCD}
\end{figure}

%
%
%
%
\vspace{-0.6cm}
\section{Results}\label{sec:results}

In Fig. \ref{fig:LCcomplete} we show both the 1.5--12 keV monitoring 15-day average light curve from the All-Sky Monitor (ASM; \citealt{levine1996}) and the 2--16 keV  light curve obtained from RXTE/PCA pointed observations of EXO 0748$-$676. The intensity of the source was approximately constant, from MJD $\sim$ 50000 to MJD $\sim$ 53000, after which a number of X-ray flares occurred, and EXO 0748$-$676 transitioned to the quiescent state. 

In Fig. \ref{fig:CCD} we show the colour-colour diagram (CCD) of EXO 0748$-$676. This plot resembles the CCD of the atoll source 4U 1636$-$53 \citep{altamirano2008b}, even if the softest branch is missing (so-called banana branch).
We identified the possible extreme island state with the hardest colors (which is the most populated part of the diagram), the island state with intermediate colors, and the lower-left banana state with the softest colors. This last identification is supported by the detection of kHz QPOs \citep{homan2000}, which are generally observed in this state (e.g., review by \citealt{vanderklis2000}). A reliable identification of the states would require a detailed variability study (e.g., \citealt{vanstraaten2003}, \citealt{altamirano2008a}) which is beyond the scope of this {\it Letter}.

We detected mHz QPOs with frequencies between $\sim$ 5.3 and $\sim$ 12.8 mHz in 14 data-segments (all shorter than 4~ksec, distributed in 11 observations) during the flaring period between 2004 March and 2007 December (MJD 53071 and 54439, respectively; see Fig. \ref{fig:LCcomplete}). All detections were found when the source was between $\sim1\%$ and $\sim3\%$ $L_{\rm Edd}$, assuming a distance of 7.1 kpc and a 3--50 keV $L_{\rm Edd} = 2.5\times10^{38}$ erg s$^{-1}$.

%

%
The mHz QPOs (filled red squares; see Fig. \ref{fig:CCD}) appear when the source have hard colour $\lesssim$ 0.9 and soft colour $\lesssim$ 1, that is, probably the lower-left banana. We also found mHz QPOs in two segments that fall in an intermediate state, close to the transition between the banana and the island states. 
In all cases the 2--5 keV fractional rms amplitude of the QPOs was $\sim$ 4\%, and the QPOs were not significantly detected for energies higher than $\sim$ 7~keV, consistent with the results of \citet{revnivtsev2001}.


\begin{figure}
\centering
\begin{tikzpicture}
\node[anchor=south west,inner sep=0] (image) at (0,0) {\includegraphics[width=\columnwidth]{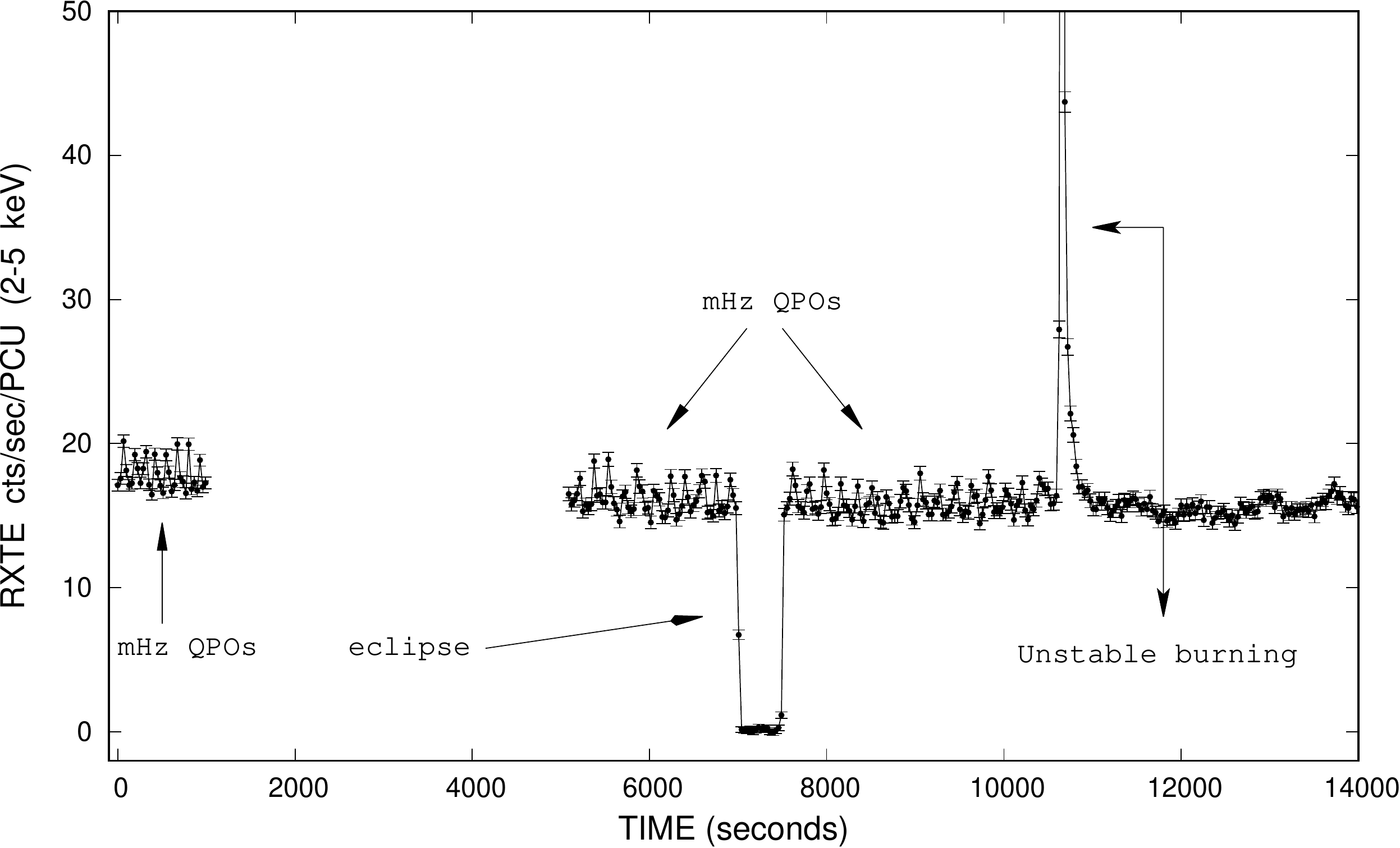}};
        \begin{scope}[x={(image.south east)},y={(image.north west)}]
            \node[anchor=south west,inner sep=0] (image) at (0.1,0.532) 
            {\includegraphics[width=0.17\textwidth]{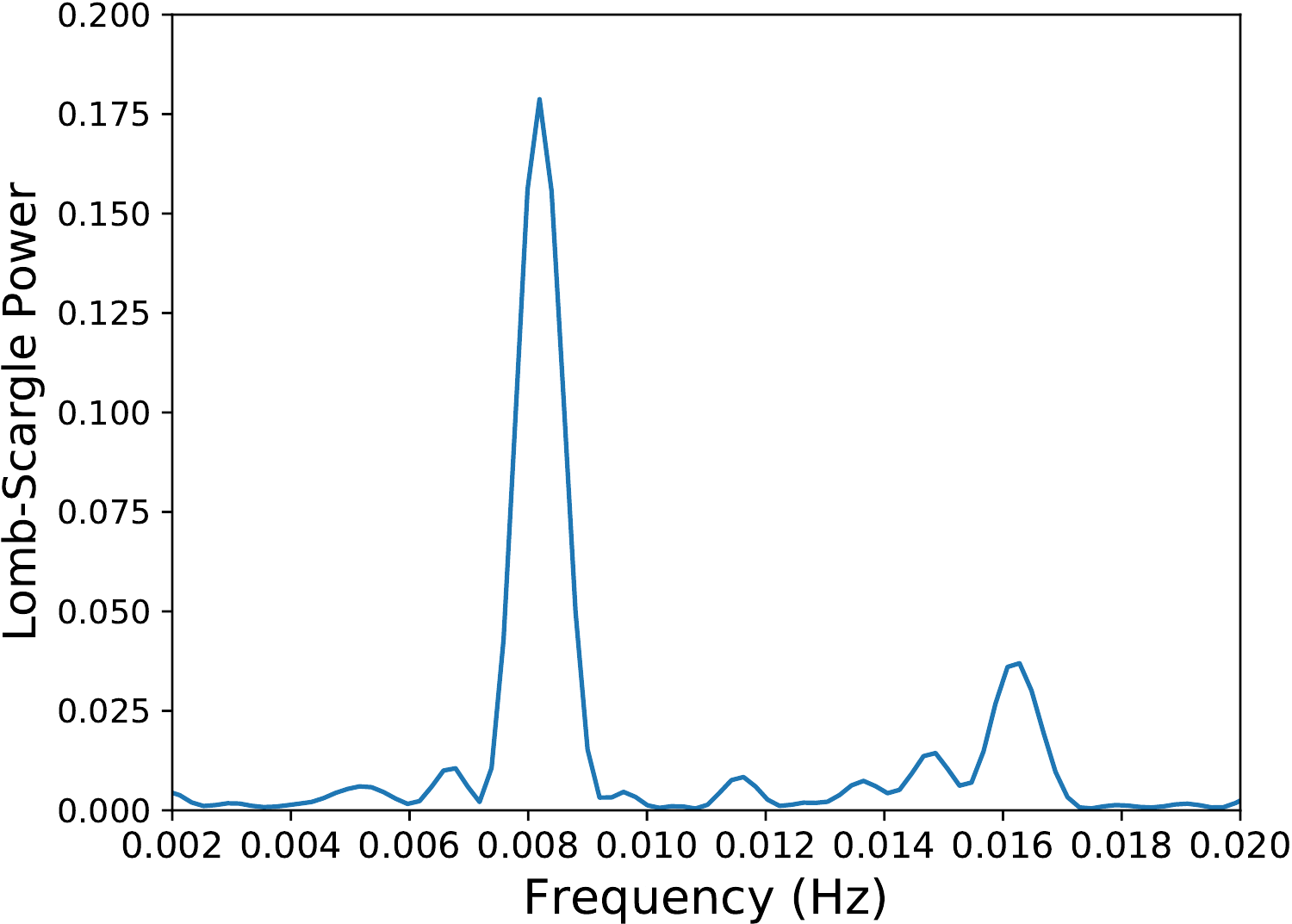}};
        \end{scope}
    \end{tikzpicture}
\caption{RXTE/PCA light curve ($\approx$ 2--5 keV) of EXO 0748--676 rebinned to 32~s in which the mHz QPOs are detected (obs ``B''; label as in Figs. 1 \& 2). 
The oscillations are present in three segments, two before and one after the eclipse. About 10.5 ks from the start of the first observation, a type I X-ray burst occurred and the QPOs, apparent before the burst, disappeared. The inset shows the Lomb-Scargle periodogram of the first $\sim$ 1200 s of the data).
A significant QPO is present at a frequency of 8 mHz, and a second harmonic component at $\sim$ 16 mHz appears in the plot.}
\label{fig:LC153-154}
\end{figure}

In Fig. \ref{fig:LC153-154} we show a representative example of a background-subtracted light curve of two consecutive observations (obsIDs 80040-01-11-00, -01; from now on, observation ``B''; see Figs. 1 \& 2). The oscillations can be seen from the beginning of the first observation to $\sim$ 10.5 ksec 
until an X-ray burst occurred, after which the QPOs disappeared.
The QPOs were not detected after the burst for 11.3 ksec until the end of that observation. The disappearance of the mHz QPOs after the X-ray burst is consistent with what \citet{revnivtsev2001} and \citet{altamirano2008b} found for other sources, and suggests that the physical processes producing the QPOs and the bursts are connected. 
In the inset of Fig. \ref{fig:LC153-154} we show the periodogram of the first data-segment. A prominent QPO at $\sim$ 8 mHz and a possible harmonic at $\sim$ 16 mHz can be seen. 

\begin{figure}
\includegraphics[height=5.3cm, width=0.95\columnwidth]{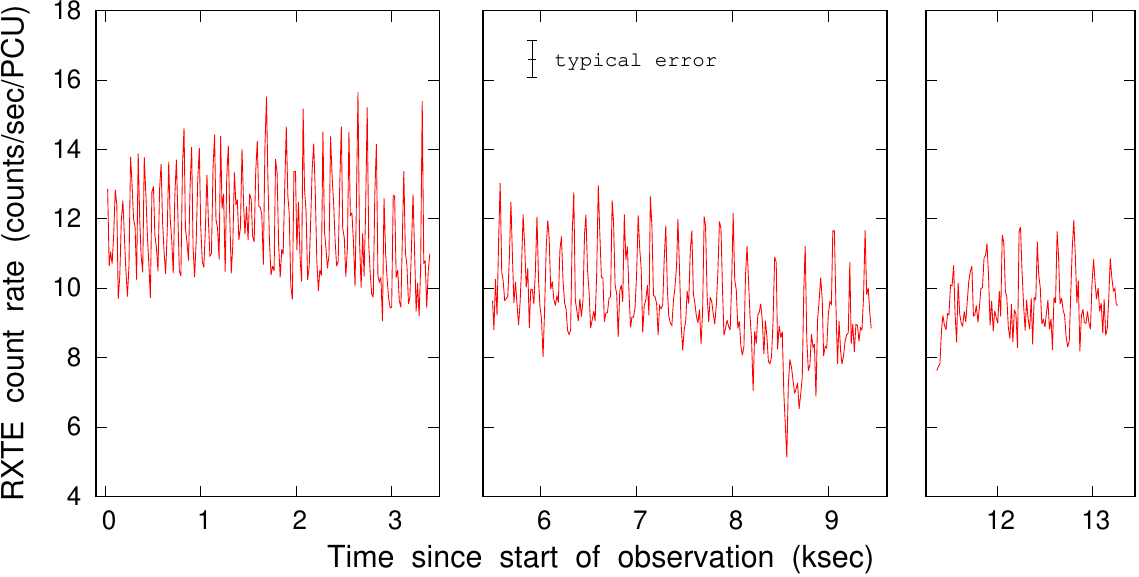}
\centering
\caption{Background-corrected light curve ($\approx$ 2--5 keV) of the observation ``A'' (label as in Figs. 1 \& 2) of EXO 0748--676 taken with RXTE/PCA, using 16-s time bins. Quasi-periodic oscillations are clearly visible through the whole observation.}
\label{fig:LC198-199-200}
\end{figure}

\begin{figure}
\includegraphics[height=5.3cm,width=0.95\columnwidth]{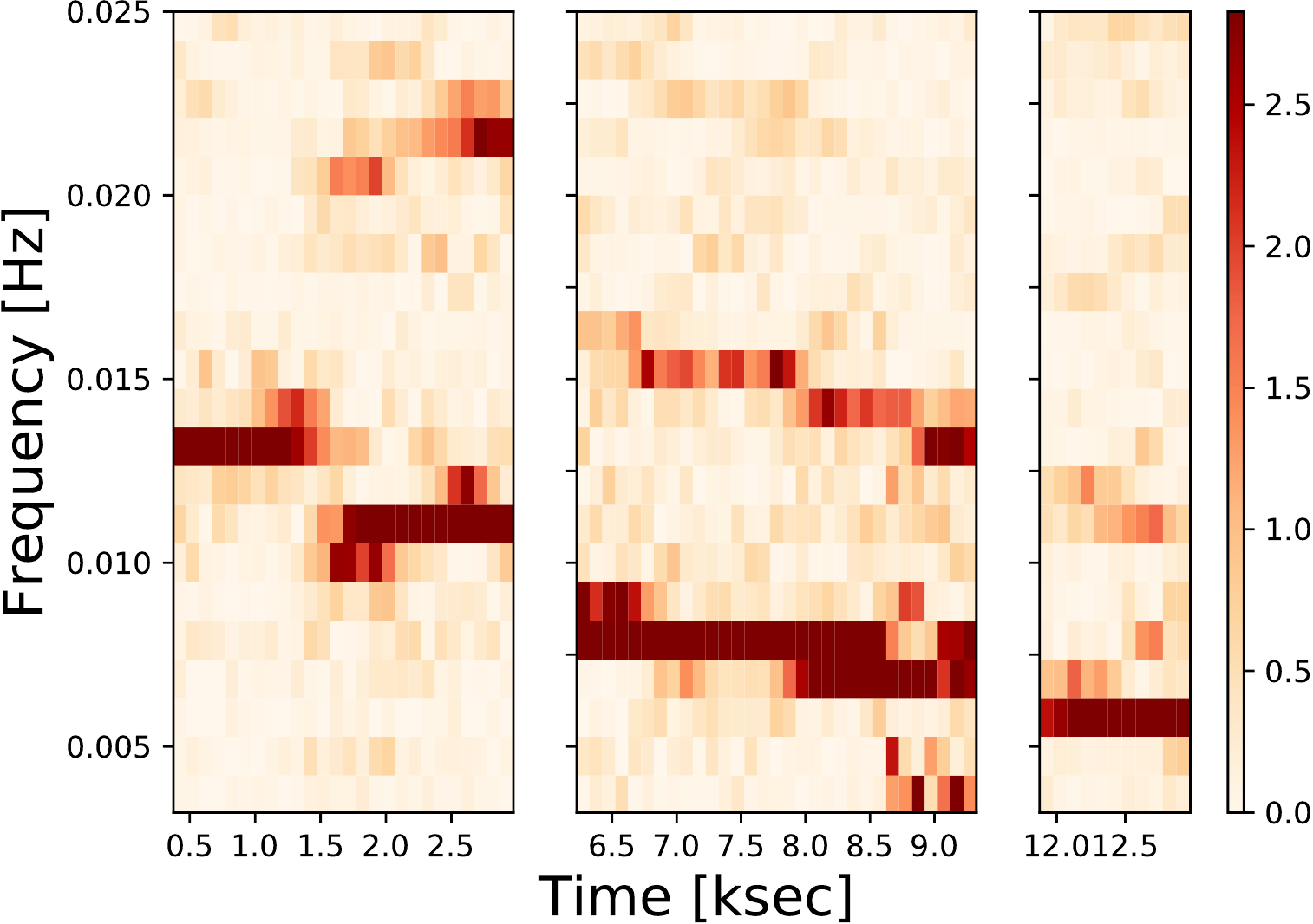}
\centering
\caption{Dynamical power spectrum corresponding to the RXTE observation ``A'' (label as in Figs. 1\& 2) of EXO 0748--676. The plot, which was made using a 700 s sliding window with steps of 50 s, shows the evolution of a frequency component associated to the mHz QPOs, decreasing from $\approx$ 13 mHz down to $\approx$ 6 mHz.}
\label{fig:Dynamical}
\end{figure}

\citet{altamirano2008b} found a systematic drift in the frequency of the mHz QPOs of some observations of the NS LMXB 4U~1636--53. 
The length of the segments where we detected mHz QPOs are generally too short to measure frequency drift; however in two cases we were able to combine data to test for drifts.
The first case is shown in Fig.~\ref{fig:LC153-154}. The QPO frequency drifts from segment to segment, starting at $\sim$ 8.2 mHz (first segment), and drifting to $\sim$~5.8 and $\sim$ 5.4 mHz, before and after the eclipse, respectively. Assuming that the QPOs were always present until the burst onset, the frequency decreased at an average rate of $\sim$~0.26 mHz ksec$^{-1}$.
The second case is from obsID 90039-01-03-05 (3 orbits; from now on, observation ``A''; see Figs. 1 \& 2), where the QPO frequency drifts from about $\sim$ 13 mHz to less than 6 mHz. Fig.~\ref{fig:LC198-199-200} shows the light curve and Fig.~\ref{fig:Dynamical} the dynamical power spectrum of this observation.
The QPOs in this observation persisted for about 13~ksec (assuming continuity during data-gaps), and their frequency drifted at an average rate of $\sim$ 0.56 mHz ksec$^{-1}$. 

EXO 0748--676 was close to the intermediate state when the QPOs exhibited the frequency drifts, consistent with what \citet{altamirano2008b} found in 4U~1636--53. Finally, excluding those cases where we observed a drift, the frequency of the mHz QPOs were between $\sim$ 6.7 and $\sim$ 8.7 mHz. Our results, therefore, show that EXO 0748--676 is the sixth source that displays this kind of QPO and the second one where a frequency drift is observed.

\vspace{-0.5cm}
\section{Discussion}

We report the discovery of mHz quasi-periodic oscillations in the NS LMXB EXO 0748--676. 
The QPO frequency ranged between $\sim$ 5 and $\sim$ 13 mHz, its amplitude was $\sim$ 4\% at low energy (2--5 keV), and in at least two cases the QPO frequency decreased in time. All the detections occurred when the source energy spectrum was relatively soft, as suggested by the colour-colour diagram. In one case we can confirm that the QPO disappeared after the onset of a thermonuclear X-ray burst (in none of the cases we saw the QPO disappear without the onset of a burst).
%
%
All these observational properties are consistent with those observed in the accreting NS LMXBs 4U 1636--53, 4U 1608--52 and Aql X--1 \citep{revnivtsev2001,altamirano2008b}, 4U 1323--619 \citep{strohmayer2011} and GS 1826--238 \citep[][although in this source the amplitude of the mHz QPO increased with energy]{strohmayer2018}, suggesting that we are observing the same phenomenology in 6 different accreting systems. These mHz QPOs have been associated with an oscillatory mode of He burning on the NS surface, generally referred as marginally stable nuclear burning \citep[e.g.,][]{heger2007,keek2009}.

\citet{linares2010} reported the detection of mHz QPOs with a frequency of $\sim$ 4.5 mHz during the 2010 outburst of the NS transient IGR J17480--2446. In later work, \citet{linares2012} reported that these mHz QPOs appear as the result of a smooth evolution of the thermonuclear bursts properties: as the luminosity of the source increased (decreased), the burst rate increased (decreased), and the fluence of each burst decreased (increased), so that at luminosities higher than $L_{2-50\hspace{0.7mm} \rm keV} \sim 10^{38}$ erg s$^{-1}$, bursts appeared in the light curve as a QPO with a period of $\sim$ 3 minutes. These results suggested that the burst rate and burst fluence are a function of the luminosity (accretion rate). This behaviour is inconsistent with our findings, where the mHz QPO and the type I X-ray bursts occur at the same observed luminosity. However, it is currently under debate if the phenomenology observed in  IGR J17480--2446 and the 6 sources listed above are produced by different physical processes, or it is the same physical process under different conditions \citep[see discussion in][]{linares2012}. For example, different composition of the burning material, different accretion geometry (i.e., accretion mainly onto the NS equator vs. spherical accretion onto most of the NS surface), different NS spin frequency and/or magnetic field, and a combination of all these, can affect the burning process and how we detect it.

\citet{ferrigno2017} detected a QPO at an approximately constant frequency of $\sim8$ mHz in the accreting millisecond X-ray pulsar (AMXP) IGR J00291+5934. 
%
%
The QPO was persistently present during an 80-ksec XMM-Newton observation. Its rms amplitude strongly decreased with energy, from $\sim30$\% at $\lesssim1$~keV to $\sim6$\% at $\sim$10~keV (the 0.5-11 keV average rms amplitude was $\sim15$\%). 
%
%
\citet{ferrigno2017} argued that the non detection of type I X-ray bursts since IGR J00291+5934 discovery (\citealt[][and also see \citealt{defalco2017}]{eckert2004,markwardt2004,falanga2005}), together with the measured properties of the oscillations, strongly suggested that the QPOs they found were not due to marginally stable burning (at least, not in the form of those described in \citealt{revnivtsev2001,heger2007,altamirano2008b}). 
\citet{ferrigno2017} instead suggested that the oscillations they found were more similar to those of the so-called ``heartbeats'' (and other type of similar variability) observed in the black holes GRS 1915+105 (see, e.g., \citealt{belloni2000}), IGR J17091-3624 \citep{altamirano2011} and potentially the NS system known as the Rapid Burster \citep{bagnoli2015}. 

\vspace{-0.5cm}
\subsection{Frequency drifts and source state}
\citet{altamirano2008b} found that the mHz QPOs in 4U 1636--53 would show clear frequency drifts if the system was observed close to the transition between the island and banana states. If the QPOs were detected in the banana state, then the QPO frequency did not drift. Our results on EXO~0748--676 are consistent with those of \citet{altamirano2008b}: in the only two instances in which we observe frequency drifts (A and B, see Fig. \ref{fig:CCD}), EXO~0748--676 was in an intermediate state.

The QPOs in observation ``A'' showed a faster frequency drift than we observed in observation ``B'' ($\sim$ 0.56 mHz ksec$^{-1}$ vs. $\sim$ 0.26 mHz ksec$^{-1}$, respectively). In terms of spectral states, observation ``A'' is at a harder state than ``B''. Under the assumption that in the banana state we would not see any frequency drift \citep[as reported by][for 4U 1636--53]{altamirano2008b}, our results appear to suggest that the rate at which the QPO frequency decreases is related to the spectral state, where the rate is maximum at the hardest colours, and decreases as the source evolves to softer states, where we do not expect to see any drift at all. This is not a scenario that can be tested with our RXTE data of EXO~0748--676; however, it could be tested with other sources (e.g., in 4U 1636--53, where \citealt{altamirano2008b} reported 22 instances of mHz QPOs with frequency drifts).

Finally, our results show that the frequency drift is not a unique characteristic of the marginally stable burning in 4U 1636--53, but that can be seen in other sources. This opens a window of opportunity to potentially better understand this type of burning. For example, comparison of the mHz QPOs characteristics between sources could shed light into the dependence of the QPO frequency drift, and the change in heat flux from the NS crust \citep{keek2009}.

\vspace{-0.5cm}
\section*{Acknowledgments}
GCM thanks the Royal Society International Exchanges program for their support. DA acknowledges support from the Royal Society. JAC, FG and GCM were partially supported by PIP 0102 (CONICET). This work received financial support from PICT-2017-2865 (ANPCyT). Lyu is supported by National Natural Science Foundation of China (grant No.11803025); and Hunan Provincial Natural Science Foundation (grant No. 2018JJ3483).

This research has made use of data and/or software provided by the High Energy Astrophysics Science Archive Research Center (HEASARC), which is a service of the Astrophysics Science Division at NASA/GSFC and the High Energy Astrophysics Division of the Smithsonian Astrophysical Observatory.This research has made use of NASA's Astrophysics Data System.

\vspace{-0.5cm}
\bibliographystyle{mnras}
\bibliography{biblio.bib}
\vspace{5mm}


\label{lastpage}
\end{document}